\documentstyle[12pt,aasms4]{article}

 \received{}

\accepted{} \journalid{}{} \articleid{}{}

\slugcomment{ to appear in the Astrophysical Journal}

\begin{document}

\def\lsim{\mathrel{\hbox{\rlap{\hbox{\lower4pt\hbox{$\sim$}}}\hbox{$<$}}}}
\def\gsim{\mathrel{\hbox{\rlap{\hbox{\lower4pt\hbox{$\sim$}}}\hbox{$>$}}}}

 \title{Discovery of Spectral Variability of  Markarian 421  at TeV Energies  }

\author{F. Krennrich\altaffilmark{1},      
I.H.~Bond\altaffilmark{2}, S.M. Bradbury\altaffilmark{2},
J.H. Buckley\altaffilmark{3},  D.A.~Carter-Lewis\altaffilmark{1},
 W.~Cui\altaffilmark{4},  
I.~de la Calle Perez\altaffilmark{2}, D.J.~Fegan\altaffilmark{5},
S.J.~Fegan\altaffilmark{6}\altaffilmark{,15} , J.P.~Finley\altaffilmark{4},
J.A.~Gaidos\altaffilmark{4}, K.~Gibbs\altaffilmark{6},
G.H.~Gillanders\altaffilmark{7},  T.A.~Hall\altaffilmark{1}\altaffilmark{,16},
A.M. Hillas\altaffilmark{2}, J.~Holder\altaffilmark{2}, 
D.~Horan\altaffilmark{6},  M.~Jordan\altaffilmark{3}, 
M.~Kertzman\altaffilmark{8}, D.~Kieda\altaffilmark{9},
J.~Kildea\altaffilmark{5}, J.~Knapp\altaffilmark{2},
K.~Kosack\altaffilmark{3},  M.J.~Lang\altaffilmark{7},
S.~LeBohec\altaffilmark{1},
P.~Moriarty\altaffilmark{10},  D. M\"uller\altaffilmark{11},
R.A.~Ong\altaffilmark{12},   R.~Pallassini\altaffilmark{2},
D.~Petry\altaffilmark{1},  J.~Quinn\altaffilmark{5},
N.W.~Reay\altaffilmark{13}, P.T. Reynolds\altaffilmark{14},           
H.J.~Rose\altaffilmark{2}, G.H. Sembroski\altaffilmark{4},  
R.~Sidwell\altaffilmark{13},  N. Stanton\altaffilmark{13},
S.P.~Swordy\altaffilmark{11}, V.V. Vassiliev\altaffilmark{9},
S.P.~Wakely\altaffilmark{11},  T.C.~Weekes\altaffilmark{6} }

\altaffiltext{1}{Department of Physics and Astronomy, Iowa State
University, Ames, IA 50011}

\altaffiltext{2}{Department of Physics, University of Leeds,
Leeds, LS2 9JT, Yorkshire, England, UK}

\altaffiltext{3}{Department of Physics, Washington University, St.~Louis,
MO 63130}

\altaffiltext{4}{Department of Physics, Purdue University, West
Lafayette, IN 47907}

\altaffiltext{5}{Physics Department, National University of Ireland,
Belfield, Dublin 4, Ireland}

\altaffiltext{6}{ Fred Lawrence Whipple Observatory, Harvard-Smithsonian
CfA, Amado, AZ 85645}

\altaffiltext{7}{Physics Department, National University of Ireland,
Galway, Ireland}

\altaffiltext{8}{Physics Department, De Pauw University, Greencastle, 
                   IN, 46135}

\altaffiltext{9}{High Energy Astrophysics Institute, University of Utah,
Salt Lake City, UT 84112}

\altaffiltext{10}{School of Science, Galway-Mayo Institute of Technology,
Galway, Ireland}

\altaffiltext{11}{Enrico Fermi Institute, University of Chicago,   Chicago, IL 60637}

\altaffiltext{12}{Department of Physics, University of California, Los Angeles, CA 90095}

\altaffiltext{13}{Department of Physics, Kansas State University, Manhattan, KS 66506}

\altaffiltext{14}{Department of Physics, Cork Institute of Technology, Cork, Ireland}

\altaffiltext{15}{Department of Physics, University of Arizona, Tucson, AZ 85721}

\altaffiltext{16}{Department of Physics \& Astronomy, University of Arkansas at 
Little Rock, AR 72204}

\clearpage
\begin{abstract}
The  detection of spectral variability of the $\gamma$-ray blazar
Mrk~421 at TeV energies is reported.  Observations with
the Whipple Observatory 10~m $\gamma$-ray telescope taken in
2000/2001 revealed exceptionally strong and
long-lasting flaring activity.  Flaring levels of 0.4 to 13 times that of the 
Crab Nebula flux provided  sufficient statistics for 
a detailed study of the energy spectrum  between 380~GeV and
 8.2~TeV as a function of flux level. These spectra are well described by a
power law with an exponential cutoff:
 $\rm \: \:  \: {{dN}\over{dE}} \propto \: E^{-\alpha}\times e^{-E/E_{0}} \:
\: m^{-2} \: s^{-1} \: TeV^{-1} $. 

There is no evidence for variation in the cutoff energy with
flux, and all spectra are consistent with an average value
for the cutoff energy of 4.3 TeV.
The spectral index varies between $\rm 1.89\pm 0.04_{stat} \pm 0.05_{syst} $ in
a high flux state and $\rm  2.72 \pm0.11_{stat} \pm 0.05_{syst} $ in a
low state.  

The correlation between spectral index and flux is tight 
when averaging over the total 2000/2001 data set. Spectral measurements
of Mrk~421 from previous years (1995/96 and 1999) by the Whipple
collaboration are consistent with this flux-spectral index correlation, 
which suggest that this  may be a constant or a 
long-term property of the source.
 If a similar flux-spectral index correlation were found for other 
$\gamma$-ray blazars, this universal property could help disentangle
the intrinsic emission mechanism from external absorption effects.

\end{abstract}

\keywords{BL Lacertae objects: individual (Markarian ~421)
--- gamma rays: observations}

\section{INTRODUCTION}

The discovery of more than 70 active galactic nuclei (AGNs) by the
EGRET $\gamma$-ray detector (Hartman et al. 1999) operating at 
$\rm E > 30$~MeV gave a fresh perspective on the AGN phenomenon, 
particularly relevant to understanding  the intrinsic 
properties of their jets.
EGRET-detected AGNs are typically radio-loud and  show a second peak in their
$\rm \nu F_{\nu}$ distribution at GeV energies. 
Blazars detected at TeV energies have a primary peak at X-ray
energies and a second component at TeV energies.
Both types are $\rm \gamma$-ray blazars and the commonly-accepted model
is that they  have their  jet oriented towards the observer revealing 
emission regions that  are strongly Doppler-boosted. 
Relativistic boosting gives rise to large flux variations
(Catanese et al. 1997) and short time scale phenomena (Gaidos
et al. 1996).   Two AGNs (Mrk~421 and Mrk~501) show emission extending
to energies greater than 10 TeV (Aharonian et al. 1999; 
Krennrich et al. 2001).

Since the discovery of TeV  $\gamma$-rays from the blazars Mrk~421 (Punch et al.
1992) and Mrk~501 (Quinn et al. 1996), these objects  played a significant role
in  discussions involving  the emission processes in AGN 
jets and attenuation effects of TeV $\gamma$-rays propagating
over extragalactic distances. Both blazars exhibit episodes of strong 
flaring activity, providing good statistics for detailed measurements of 
their average energy spectra from 260~GeV up 17~TeV using ground-based
$\gamma$-ray telescopes. Mrk~421 and Mrk~501 are
at approximately the same distance (z=0.031  and z=0.034, respectively).
Since  the level of attenuation of $\gamma$-rays by the diffuse extragalactic 
background light (EBL) via pair creation (Nikishov 1962; Gould \& Schr\`eder 1967;
Stecker, De Jager \& Salamon 1992)  depends on the distance of the source to the observer, 
it could cause a common spectral feature in the energy spectra of  Mrk~421 and Mrk~501.

Measurements by the Whipple collaboration (Samuelson et al. 1998; 
Krennrich et al. 2001) imply that the energy spectra of 
both Mrk~501 and Mrk~421 require a curved fit parametrization,
e.g., a power law with an exponential cutoff  with cutoff 
energies of $\rm 4.6 \pm 0.8_{stat}$~TeV   and $\rm
4.3 \pm 0.3_{stat} (-1.4 +1.7)_{syst}$~TeV (``stat'' means statistical error,
``syst'' means systematic error), respectively. Data from the HEGRA
collaboration suggest that the cutoff energy of Mrk~501 is 
$\rm 6.2 \pm 0.4_{stat} (-1.5 +2.9)_{syst} $~TeV (Aharonian et al.
1999; Aharonian et al. 2001) and that Mrk~421 has a cutoff energy
of $\rm 4.2 \:  (+0.6 -0.4)_{stat}$~TeV (Kohnle et al. 2001).   
The results from  the two  groups may be consistent given 
systematic uncertainties.  The interpretation of the origin of the 
cutoff requires a better understanding of the emission process in 
$\rm \gamma$-ray blazars as a class of objects. 
Making progress in unraveling the emission process in blazars
from  external attenuation effects requires two types of key observations.

The first is contemporaneous multiwavelength observations at X-ray and TeV 
energies (see Buckley et al. 1996; Krawczynski et al. 2002). These 
constrain the emission process by simultaneously probing the synchrotron 
emission and the mechanism responsible for the $\rm \gamma$-ray  component,
 e.g., the inverse Compton scattering process (Maraschi, Ghisellini \& Celotti 1992;
Marscher \& Travis 1996). There are also competing models that assume that the
high energy emission arises from protons (Mannheim 1998; 
M\"ucke \& Protheroe 2001; Aharonian 2000).

The second is the study of $\gamma$-ray spectral variability as a 
function of flux and time.  Spectral variability is directly tied 
to the emission process at  the source, whereas external absorption 
by the EBL is a universal feature that is independent of flux level.  
Thus, by studying the  spectral variability, we are able to disentangle 
these two effects.   In  synchrotron-Compton models,
   spectral variability could be explained by cooling of 
   electrons in the jet causing a shift of the break in the
   electron spectrum, or by variations of the maximum energy of
   accelerated electrons.  Spectral variability may also be a  key to
   the understanding of the energy dissipation processes in the vicinity of a
   supermassive black hole powering the jet.

Some evidence for spectral variability has been reported by
Djannati-Ata\"{\i} et al. (1999) and Krawczynski et al. (2001) for
Mrk~501; however, the effect was not highly statistically significant 
($\rm \approx$~4 standard deviations)  and precluded detailed studies. 
In fact, the spectral index of Mrk~501 turned out to be 
surprisingly stable during observations of a strong outburst in 1997
(Aharonian et al. 1999) though the flux varied by a factor of 30.

In this paper we present the discovery of spectral variability
 of Mrk~421 at TeV energies based on data
taken with the Whipple Observatory 10~m $\rm \gamma$-ray
telescope. We show that Mrk~421 exhibits a remarkable
flux-spectral index correlation that appears to
be stable  averaged over time-scales  of months to several years. 

\section{OBSERVATIONS \& DATA ANALYSIS}

The observations were made with the Whipple Observatory 10~m $\rm
\gamma$-ray telescope using the GRANITE-III high-resolution camera
(Finley et al. 2001). The high sensitivity of the telescope for point sources
 in the energy range from $\rm \approx$~200~GeV to greater than
20~TeV  permits the measurement of energy spectra of $\gamma$-ray sources
with fluxes above 1~Crab  on time-scales as short as 30~minutes 
(1~Crab is defined  as the differential flux
 of the Crab Nebula at 1~TeV in $\rm photons \:  m^{-2} \: s^{-1} \:  TeV^{-1}  $). 

Mrk~421 was more active in the 2000/2001 observing season than in previous years. 
Unusually high flaring states ranging from 0.4 Crab to 13~Crab,
lasting for a period of approximately five months, provided a wealth of statistics
for studies of spectral variability.
The data used for the  analysis in this paper were collected on 2000
November 28, December 4-6, 22, 27, 28-29 and 2001 January 20-21, 24-25, 30-31,
February 1-3, 19, 27-29, March 19, 22-23, 26-27, 30 and April 13.
These data  are an extension to the data set published in Krennrich et al. (2001),
and include observations of the source in a lower flux state.
A total of 49.93 hours of on-source observation time with  zenith angles less than
$\rm \approx 35^{\circ}$  has been used in this study.

The $\gamma$-ray rate of the 107 individual on-source runs varies from 0.4
to 18.0~$\rm \gamma \: minute^{-1}$.  The background from cosmic-ray
induced showers has been estimated  for each on-source run
individually by using a matching off-source run also taken
during the 2000/2001 season.  A good match requires 
that both  runs cover a similar zenith angle range, and that the
 on-source and off-source runs show good agreement in the distribution of the
  parameter associated with the alignment of the image in the focal
  plane, for values of the parameter outside the $\rm \gamma$-ray fiducial
  region.     In some runs, a normalization factor between
  the on and off samples was applied to ensure that the off-samples
  accurately represented the background in the on-region.
This procedure has  been tested as a function of the total light intensity 
of  the $\gamma$-ray image to minimize a possible systematic bias.  Uncertainties 
in the spectral index, alpha,  due to the method of background 
estimation are typically $\rm \Delta \alpha < 0.1$ in the spectral index for 
individual runs  and $\rm \Delta \alpha <0.05$ for sets of 5 or more runs.

In a search for yearly trends in a flux-spectral index correlation we also include previously
published data from 1995 and 1996 (Krennrich et al. 1999a). In addition we
derive a spectrum for observations taken in May-June 1999
using the GRANITE-III 331-pixel camera (Krennrich et al. 1999b; Le Bohec et al. 2000;
Finley et al. 2001).     The data consist of 33 on-source and 33 
matched off-source runs from 1999 May 6-7, 9-10, 11, 16-18 and  1999 June 5-8.

The analysis methods for the 2000/2001 observations,  $\gamma$-ray selection and 
energy estimate  are based on the  method described in Mohanty et al. (1998)
and their application has been described in Krennrich et al. (2001).    
These $\gamma$-ray selection criteria are derived from parameter distributions of simulated
$\gamma$-ray showers as a function of their total light intensity in the camera.
We set these criteria so that they retain 90\% of $\gamma$-ray images whose
centroid is within  $\rm 0.4^{\circ} - 1.0^{\circ}$ from the center of the camera.
To avoid the difficulties of modeling the trigger electronics we apply an additional cut
requiring that a signal of at least 15.1,  13.6,  and 12.1 photoelectrons are present
in the three highest camera pixels, respectively.    
In this analysis, we have increased the
lowest energy point of our spectral measurements from 260 GeV to 380 GeV, to minimize
 the systematic uncertainties inherent at low energies.

\section{RESULTS: SPECTRAL VARIABILITY AS A  FUNCTION OF FLUX}

The data were divided a priori into eight independent subsets  with
comparable numbers of excess events  and average $\rm \gamma$-ray rates ranging from 
3.3 to 16.0~$\rm \gamma \: minute^{-1}$.
In Figure~1 we show the corresponding energy  spectra. For clarity of 
presentation  we have combined set II and III,  and set VI and VII, respectively. 
Progressive hardening of the spectra is apparent to the eye when comparing  the spectra 
towards increasing flux levels.
We have attempted to fit the individual  spectra  by a simple power law.
These fits result in  unacceptable  goodness of fit ($\rm \chi^{2}$) values for 
sets I-IV,  hence the power law hypothesis is rejected   (also see Table 1).

For comparison with previous papers (Krennrich et al. 1999a; 
Krennrich et al. 2001)
 and the results from other groups (Piron et al. 2001; Bazer-Bachi et al. 2001), we also
fit the data using a parabolic function:
$ \rm    {{dN}\over{dE}}~\propto~$~$\rm~E^{-\alpha
 \: -\beta log_{10}(E) } \:
\: m^{-2} \: s^{-1} \: TeV^{-1} $.
The results in Table 2 have acceptable goodness of fit ($\rm \chi^{2}$) 
values.   The spectrum hardens with increasing flux,  whereas the curvature term 
shows no significant dependence on flux.
In a previous paper (Krennrich et al. 2001) describing the average energy spectrum of 
Mrk~421 in a high flaring state in 2001,  the best fit to the energy spectrum
was achieved by using a power law with an exponential cutoff.

\smallskip

(1) $ \rm  \:   \:  \:  {{dN}\over{dE}}  \propto    \:$  $\rm  E^{-\alpha}  \times  e^{-E/E_{0}} \:
\: m^{-2} \: s^{-1} \: TeV^{-1} $

\noindent The results for fits of this form are provided in Table~3 and exhibit
acceptable goodness of fit values for all spectra.
Since the spectral index and cutoff energy $\rm E_0$ are correlated, we 
also present the  uncertainty of  $\rm E_0$ when accounting  for  this 
correlation as shown in parenthesis in Table~3.
 These uncertainties are the  extrema of $\rm E_0$ of the 1~$\rm \sigma$
error ellipse\footnote[1]{as calculated by MINUIT, Version 94.1,
Cern Program Library entry D506.  We also compared MINUIT with
the method given by Avni et al. (1976) giving consistent results.}
that result from plotting  the minimizing function
$\rm \chi^{2}$ as a function of spectral index $\rm \alpha$ versus
cutoff energy $\rm E_0$.

Figure 2 shows the  derived cutoff energies for the individual
data sets (I-VIII) at different  flux levels in units of  Crab.   
No evidence for variability in the cutoff
energy is suggested by the data  (probability for the hypothesis
of a fixed cutoff energy is $\rm P \: = \: 0.98$).
It is, however, important to realize that the statistical uncertainties 
in the cutoff energy are strongly correlated with the spectral index and 
the error bars on $\rm E_0$ are typically of magnitude 1-3 TeV.  Therefore, we 
cannot exclude variability in the cutoff energy at the few TeV level.  Instruments 
that extend to significantly lower energies  and provide better sensitivity at 
higher energies are required to reduce the uncertainties in the cutoff energy.
In addition, our measurement of the cutoff energy has  systematic uncertainties,
e.g., the absolute energy scale,  which  can also be improved with the next generation 
of $\gamma$-ray instruments, the VERITAS project (Weekes et al. 2002).

  Hence, for the remainder of this paper              we adopt
a parametrization (see equation 2) with a fixed  cutoff energy of $\rm E_0 = 4.3~TeV$,  
the same cutoff energy that was derived for the average spectrum of Mrk~421 in a high 
flaring state (Krennrich et al. 2001).

(2) $ \rm  \:   \:  \:   {{dN}\over{dE}}  \propto    \:$  $\rm  E^{-\alpha}  \times  e^{-E/E_{0}} \:
\: m^{-2} \: s^{-1} \: TeV^{-1} $
with $\rm E_{0} = 4.3 \: \pm 0.3_{stat} \: (-1.4 \: +1.7)_{syst} \: TeV$

As can be seen from Table 3, the spectra corresponding to
different flux levels are well fit by this parametrization,
but show significant variation in the spectral index $\rm
\alpha$.   We find     that the energy spectra exhibit spectral
variability from $\rm \alpha \: = \: 1.89 \pm 0.04_{stat} \pm
0.05_{syst}$ up to $\rm  \alpha \: = \:  2.72 \pm 0.11_{stat}
\pm0.05_{syst}$.  
The systematic uncertainties, which are indicated by the shaded 
areas in Figure 1,   have been derived as in Krennrich et al. (2001) by
varying cut efficiencies, the angular resolution cut and 
the software trigger threshold.  In addition,   uncertainties due to the 
method of background matching are included.  Other potential sources of
 systematic uncertainties  such as   throughput variations (light throughput of the
atmosphere, mirror reflectivity and gain variations of the telescope)
(see LeBohec  \& Holder 2002, in preparation), elevation dependence and varying 
signal-to-background ratio  have been studied.  No systematic dependence
of the spectral index with these variables were found.

Figure 3 shows the spectral index as a function 
of the absolute $\rm \gamma$-ray flux in units of the Crab.  Variability of the 
spectral index is apparent, and therefore  the hypothesis that the spectral 
index  is constant  can be excluded at the $\rm P \: = \: 9.5 \times 10^{-14}$ level.    
The correlation between spectral index and flux indicates  substantial spectral hardening with
increasing flux. If we fit the spectral index as a function of flux
$ \rm \Phi  $ in units of Crab with a  second order polynomial we get

$\rm \alpha(\Phi) \: = \:
 -2.66 \:  (\pm 0.01)\: + \: 0.123\:  (\pm 0.030)\:
\times\: [\Phi/Crab]\:  - \:  0.0056 \: (\pm 0.00230) \:
\times \:   [\Phi/Crab]^2  $

which provides a  $\rm \chi^{2} \: = \: 13.92$ for 5 degrees of freedom
($\rm P \: = \: 1.61 \times 10^{-2}$).

\section{DISCUSSION}
Strong and extended flaring of Mrk 421 allows us to study
the spectral variability of this source as a function of flux.
Averaged over the time period         2000 November 28  to  2001 April 13  
the data show a clear flux-spectral index correlation, 
with the spectral  index varying between $\rm 1.89\pm 0.04$  in a high state and
$\rm  2.72 \pm0.11$  in a low state.

Whether or not this correlation is maintained in different
epochs outside of the 2000/2001 observing period can be addressed   using previously 
published results and  archival data from the Whipple collaboration.   Figure 3 
also shows results  from the average spectrum in 1995/1996 in a high state 
(Krennrich et al. 1999a) and  the average spectrum  from 1999 May 6 through June 8.  
 The  data points  from 
1995/1996   and from 1999  fall  into place with the flux-spectral index  correlation  
as observed for the  2000/2001 data alone.
 This indicates that  the   correlation between spectral index and flux  holds true 
when averaging over time scales of months to five years.

Spectral hardening during flares has also been observed for Mrk~421 in X-rays by 
Fossati et al. (2000)  using BeppoSAX data during X-ray flares in 1997 and 1998.  In  X-rays,
the  effect of spectral hardening  has been interpreted as a shift of the  synchrotron
 peak towards higher frequencies.    The flux-spectral index correlation seen in the TeV 
spectra  could also be 
interpreted as  a shift of the high energy peak towards higher frequencies.   
The shape of the spectrum~I in Figure 1 suggests that the peak in  $\rm \nu F_{\nu}$ is at a few 
hundred~GeV, significantly above previous levels (Maraschi et al. 1999).
The spectral hardening is most evident at energies below 2~TeV; it is not uniform.  

Strong spectral hardening at lower energies might be expected
in the inverse-Compton (IC) scenario in which the IC peak shifts
towards higher energies as the flux increases.  Conversely,
at the higher energies, spectral softening occurs due to either
a terminating particle distribution in energy,  the  falling cross-section due to the 
Klein-Nishina effect, or  external attenuation effects from  the EBL or nearby radiation 
fields.    If the changes are due to a shifting IC peak energy, the flux value would be 
closely tied to the spectral index, as seen here.
Further studies of spectral variability on short time-scales (hours-days)
will be presented in a follow-up paper.

\acknowledgments

We acknowledge the technical assistance of K. Harris, J. Melnick and E. Roache.
This research is supported by grants from the U.S. Department of Energy
and by NASA, NSF, and by PPARC in the UK and by Enterprise-Ireland.

\vfill\eject

\clearpage

\begin{center}
  \mbox{\epsfysize=0.80\textheight\epsfbox{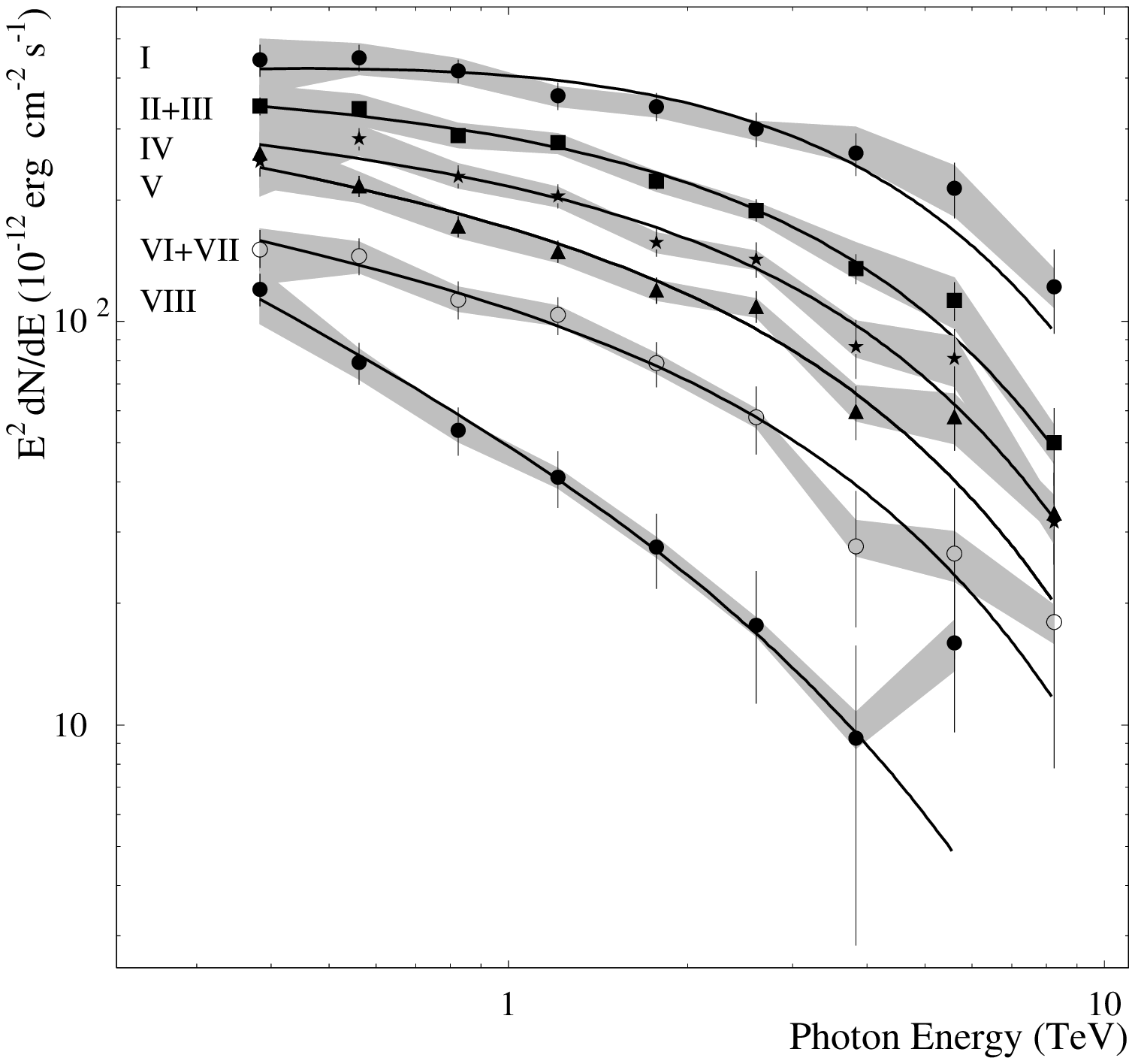}}
\end{center}

\figcaption[m42d]
{Mrk~421 spectra at different flux levels averaged for data over the 2000/2001 season.  
The spectra have been fit by a power law with a fixed exponential cutoff at 4.3 TeV 
(Krennrich et al. 2001).    The fits produce acceptable goodness of fit ($\rm \chi^{2}$)
values.  The shaded areas indicate the systematic errors on the
flux measurements.}

\clearpage

\begin{center}
  \mbox{\epsfysize=0.80\textheight\epsfbox{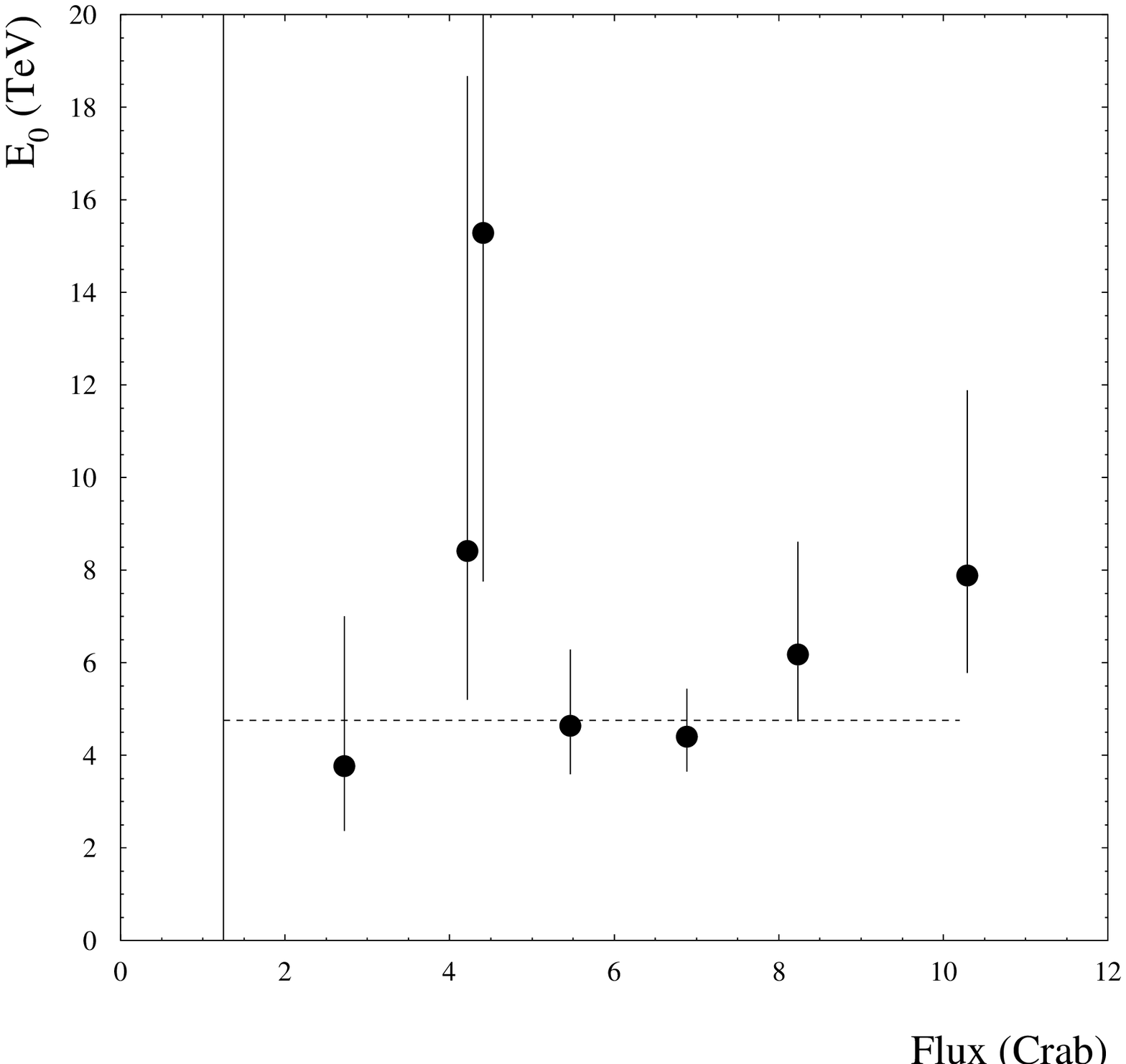}}
\end{center}

\figcaption[m42d]
{The cutoff energy $\rm E_0$ is plotted as a function of  flux in units of Crab (defined in text) for various 
flaring states (data sets I-VIII) of the 2000/2001 season.
 No significant variability of the cutoff energy is seen in the data.  Note that the point at the lowest
flux level has a large statistical uncertainty so that it is off the plot, only showing its error bar.  }

\clearpage

\begin{center}
  \mbox{\epsfysize=0.80\textheight\epsfbox{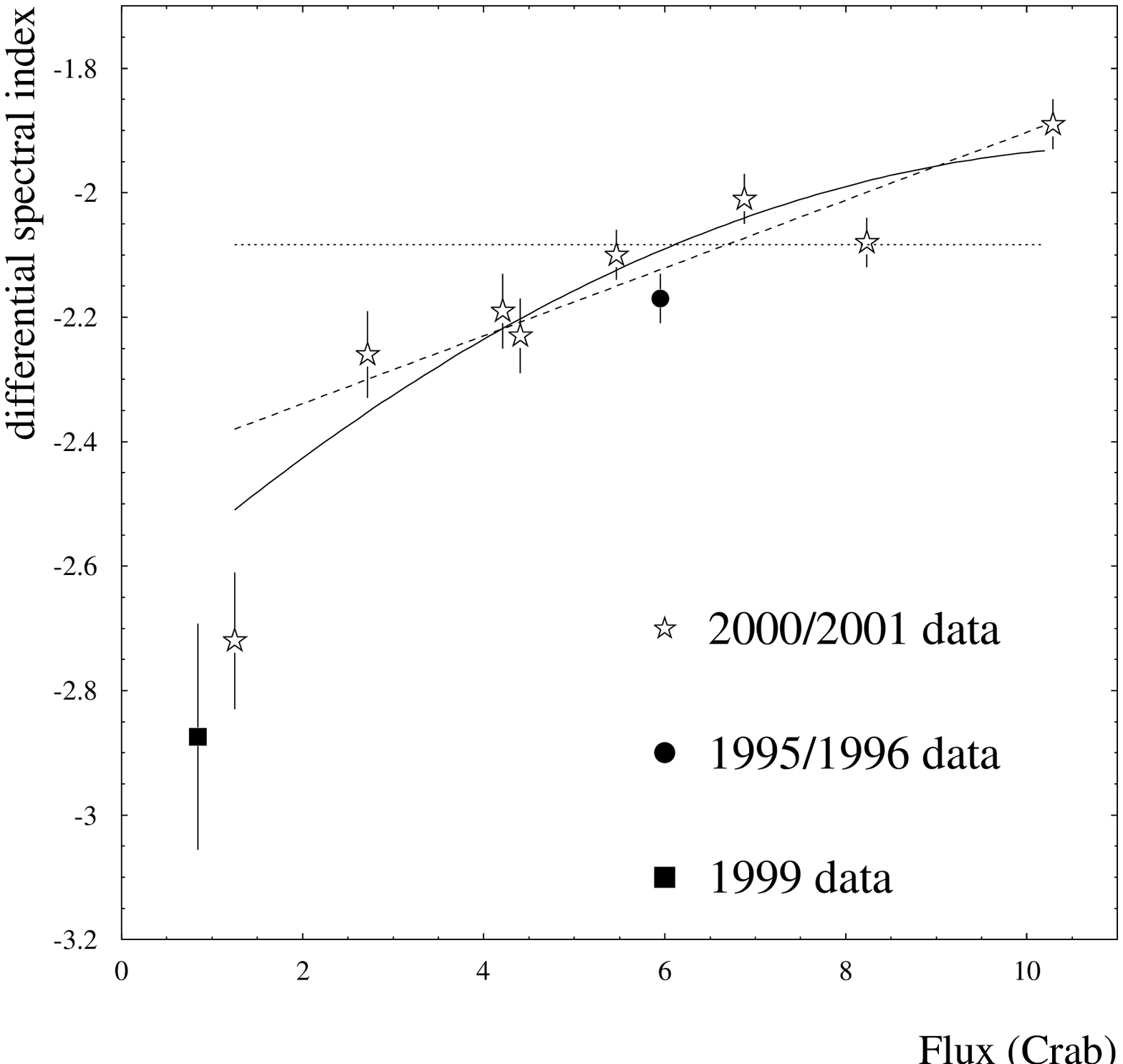}}
\end{center}

\figcaption[m42d]
{The stars show the spectral index of Mrk 421 as function of flux
(in Crab units) for the 2000-2001 data set.  The hypothesis of a
constant spectral index (dotted line) is rejected, whereas the data
are better fit by a linear relation  ($\rm P \: = \: 5 \times 10^{-3}$).
A second order polynomial (solid line) gives a better but still marginal fit  
($\rm P \: = \: 1.6 \times 10^{-2}$).  In addition,  we show results for Whipple 
1995/96 data (solid circle) from Krennrich et al. (1999a) and data taken during 1999 May-June  
(solid rectangle). }

\clearpage

\begin{deluxetable}{lcccccc}
 \footnotesize
 \tablecaption{Fits to the spectra of Mrk~421: power law$^a$
 \label{tbl-1}}
 \tablewidth{0pt}
 \tablehead{ \colhead{set} &
  \colhead{$\rm \alpha$} & \colhead{$\rm \chi^2/dof$$^b$}  \\
  }
  \startdata
I &$\rm 2.31\pm 0.04$&$\rm 1.57$\nl
II &$\rm 2.47\pm 0.03$&$\rm 3.70$\nl
III &$\rm 2.43\pm 0.03$&$\rm 5.47$\nl
IV &$\rm 2.48\pm 0.06$&$\rm 3.86$\nl
V &$\rm 2.56\pm 0.05$&$\rm 0.91$\nl
VI &$\rm 2.57\pm 0.04$&$\rm 0.82$\nl
VII &$\rm 2.60\pm 0.06$&$\rm 1.28$\nl
VIII &$\rm 2.95\pm 0.10$&$\rm 0.32$\nl
\\
\hline
 \multicolumn{4}{l}{\footnotesize
$^a$    $\rm dN/dE \propto E^{-\alpha} $  (m$^{-2}$ s$^{-1}$ TeV$^{-1}$)       }\\
\multicolumn{4}{l}{\footnotesize
$^b$ Results from the fits for the energy  }\\
 \multicolumn{4}{l}{\footnotesize spectra of Mrk 421, using a power law.  }\\
 \enddata
 \end{deluxetable}

\clearpage

\begin{deluxetable}{lccccccc}
 \footnotesize
 \tablecaption{Fits to the spectra of Mrk~421: power law
with curvature.
 \label{tbl-1}}
 \tablewidth{0pt}
 \tablehead{\colhead{set} &
 \colhead{parabolic$^a$}  & \colhead{} & \colhead{}    \\
 \colhead{} & 
 \colhead{$\rm \alpha$} & \colhead{$\rm \beta$} & \colhead{$\rm \chi^2$/dof$^b$} }
  \startdata
I &$\rm 2.22\:\pm 0.05$&$\rm 0.27\:\pm 0.10$&0.40\nl
II &$\rm 2.39\:\pm 0.04$&$\rm 0.34\:\pm 0.09$&1.53\nl
III &$\rm 2.29\:\pm 0.04$&$\rm 0.45\:\pm 0.09$&0.91\nl
IV &$\rm 2.38\:\pm 0.05$&$\rm 0.45\:\pm 0.11$&1.08\nl
V &$\rm 2.51\:\pm 0.06$&$\rm 0.20\:\pm 0.13$&0.63\nl
VI &$\rm 2.55\:\pm 0.05$&$\rm 0.13\:\pm 0.12$&0.75\nl
VII &$\rm 2.53\:\pm 0.08$&$\rm 0.49\:\pm 0.21$&0.32\nl
VIII &$\rm 2.95\:\pm 0.09$&$\rm -0.16\:\pm 0.29$&0.32\nl
\\
\hline
 \multicolumn{4}{l}{\footnotesize
$^a$    $\rm dN/dE \propto E^{-\alpha - \beta log_{10}E}$  (m$^{-2}$ s$^{-1}$ TeV$^{-1}$)       }\\
\multicolumn{4}{l}{\footnotesize
$^b$ Results from the fits
using a power law fit including curvature.  }\\
 \enddata
 \end{deluxetable}

\clearpage

\begin{deluxetable}{lccccccc}
 \footnotesize
 \tablecaption{Fits to energy spectra of Mrk~421 at various flux levels in 2001: power law including
 exponential cutoff.
 \label{tbl-1}}
 \tablewidth{0pt}
 \tablehead{\colhead{set} &
   \colhead{power law exp. cutoff$^a$}  & \colhead{} & \colhead{}  &\colhead{exp. cutoff (4.3 TeV)$^b$}   \\
 \colhead{} & 
 \colhead{$\rm \alpha$} & \colhead{$\rm E_{0}$$^c$} & \colhead{$\rm \chi^2$/dof$^d$} &\colhead{$\rm \alpha$} 
& \colhead{$\rm \chi^2$/dof$^e$}  }
  \startdata
I &$\rm 2.07\:\pm 0.09$&$\rm 7.89\:\pm 2.65\:(^{+4.1}_{-2.1})$&0.23&$\rm 1.89\:\pm 0.04$&1.04\nl
II &$\rm 2.20\:\pm 0.08$&$\rm 6.18\:\pm 1.76\:(^{+2.4}_{-1.5}$)&1.75&$\rm 2.08\:\pm 0.04$&2.13\nl
III &$\rm 2.02\:\pm 0.08$&$\rm 4.40\:\pm 0.86\:(^{+1.0}_{-0.8})$&0.30&$\rm 2.01\:\pm 0.04$&0.34\nl
IV &$\rm 2.12\:\pm 0.09$&$\rm 4.64\:\pm 1.27\:(^{+1.7}_{-1.1})$&1.28&$\rm 2.10\:\pm 0.04$&1.30\nl
V &$\rm 2.36\:\pm 0.11$&$\rm 8.41\:\pm 4.61\:(^{+10.3}_{-3.2})$&0.46&$\rm 2.19\:\pm 0.06$&0.83\nl
VI &$\rm 2.46\:\pm 0.09$&$\rm 15.29\:\pm 12.26\:(^{+100.4}_{-7.5})$&0.71&$\rm 2.23\:\pm 0.06$&1.69\nl
VII &$\rm 2.22\:\pm 0.19$&$\rm 3.77\:\pm 2.02\:(^{+3.2}_{-1.4})$&0.43&$\rm 2.26\:\pm 0.07$&0.44\nl
VIII &$\rm 2.95\:\pm 0.10$&$\rm 25,977\:\pm 84,528\:(^{+10,712}_{-50,301})$&0.38&$\rm 2.72\:\pm 0.11$&0.81\nl
\\
\hline
 \multicolumn{4}{l}{\footnotesize
 $^a$ $\rm dN/dE \propto E^{-\alpha} \: e^{-E/E_0} $
  (m$^{-2}$ s$^{-1}$ TeV$^{-1}$)}\\
 \multicolumn{4}{l}{\footnotesize
 $^b$ $\rm dN/dE \propto E^{-\alpha} \: e^{-E/E_0} $  (m$^{-2}$ s$^{-1}$ TeV$^{-1}$);
    $\rm E_0 = 4.3$~TeV  fixed}\\
\multicolumn{4}{l}{\footnotesize
$^c$ cutoff energy $\rm E_0$ (TeV) }\\
\multicolumn{4}{l}{\footnotesize
$^{d,}$$^e$  Results from the fits 
using a power law fit with an exponential cutoff.  }\\
 \enddata
 \end{deluxetable}


\begin{thebibliography}{}

\bibitem[Aharonian et al.\ 1999]{aha99}
 Aharonian, F.A., et al. 1999, A\&A, 349, 11

\bibitem[Aharonian \ 2000]{aharon00} Aharonian, F. A.
2000,  New Astronomy, 5, 377

\bibitem[Aharonian et al.\ 2001]{aha01}
 Aharonian, F.A., et al. 2001, A\&A, 366, 62

\bibitem[Avni et al.\ 1976]{kre76}
Avni, Y. 1976, \apj, 210, 642


\bibitem[Bazer-Bachi et al.\ 2001]{baz01}
Bazer-Bachi, R., et al. 2001, in AIP Conf. Proc. 558, High Energy
Gamma-Ray Astronomy, ed. F.A. Aharonian \& H.J. V\"olk (New
York:AIP), 643

\bibitem[Buckley et al.\ 1996]{buck96}
Buckley, J.H., et al. 1996, \apj, 472, L9




\bibitem[Catanese et al.\ 1997]{catan97}
Catanese, M.A., et al. 1997, \apj, 487, L143


\bibitem[Djannati-Ata\"{\i}, A.,  et al.\ 1999]{Dja99}
Djannati-Ata\"{\i}, A.,  et al. 1999, A\&A, 350, 17


\bibitem[Finley et al.\ 2001]{kre76}
Finley, J.P., et al.  2001, Proc. 27th Int. Cosmic-Ray Conf.
(Hamburg), 2827

\bibitem[Fossati et al.\ 2000]{Foss00}
Fossati, G., et al. 2000, \apj, 541, 166



\bibitem[Gaidos et al.\ 1996]{gai96} Gaidos, J. A., et al. 1996,
 \nat, 383, 319


\bibitem[Gould \& Schr\` eder\ 1967]{gould67} Gould, R. J., \& Schr\` eder, 
G. 1967, Phys. Rev., 155, 1408



\bibitem[Hartman et al.\ 1999]{hart99}
Hartman, R.C., et al. 1999, ApJS, 123, 79


\bibitem[Kohnle et al.\ 2001]{kohn01}
Kohnle, A., et al. 2001,  Proc. of 27th Int. Cosmic-Ray Conf. (Hamburg), 2605



\bibitem[Kraw \ 2000]{Kraw00} Krawczynski, H., et al. 2001, 
High Energy Gamma-Ray Astronomy, AIP Conf. Proc., 558
eds. F.A. Aharonian \& H.J. V\"olk, 639


\bibitem[Krawczynski et al.\ 2002]{kraw02}
Krawczynski, H., et al. 2002, M.N.R.A.S., in press


\bibitem[Krennrich et al.\ 1999a]{kre99}
Krennrich, F., et al. 1999a, \apj, 511, 149


\bibitem[Krennrich et al.\ 1999b]{kre99b}
Krennrich, F., et al. 1999b, Proc. of 26th Int. Cosmic-Ray Conf. (Salt Lake City),
3, 301


\bibitem[Krennrich et al.\ 2000]{kre00}
Krennrich, F., et al. 2001, \apj, 560, L45



\bibitem[Le Bohec et al.\ 2000]{bolec00}
LeBohec, S., et al. 2000, \apj, 539, 209



\bibitem[Mannheim \ 1998]{mann98}Mannheim, K. 1998,
Science, 279, 684


\bibitem[Maraschi, Ghisellini, \& Celotti \ 1992]{mann93} Maraschi, L., 
Ghisellini, G., \& Celotti, A. 1992, ApJ, 397, L5

\bibitem[Maraschi \ 1999]{mara99} 
Maraschi, L., et al.  1999, ApJ, 526, L81



\bibitem[Marscher, \& Travis\ 1996]{mann93} Marscher, A.P., \& Travis, 
J.P. 1996, A\&AS, 120, 537

\bibitem[Mohanty et al.\ 1998]{mohanty98}
Mohanty, G., et al. 1998, Astropart. Phys., 9, 15



\bibitem[M\" ucke \& Protheroe \ 2000]{muecke2000}
M\" ucke, A.,  \& Protheroe, R.J. 2000,  Astropart. Phys., 15, 121



\bibitem[Nikishov \ 1962]{niki98}
Nikishov, A.I. 1962, Soviet Phys.---JETP Lett., 14, 393


\bibitem[Piron et al.\ 2001]{Punch92}
Piron, F., et al. 2001, A\&A, 374, 895


\bibitem[Punch et al.\ 1992]{Punch92}
Punch, M., et al. 1992, Nature, 358, 477

\bibitem[Quinn et al.\ 1996]{QU96}
Quinn, J., et al. 1996, \apj, 456, L83


\bibitem[Samuelson et al.\ 1998]{sam98}
Samuelson, F. W., et al. 1998, \apj, 501, L17


\bibitem[Stecker, De Jager \& Salamon \ 1992]{stdj97}Stecker, F. W.,
De Jager, O. C., \&  Salamon, M. H.  1992, \apj, 390, L49


\bibitem[Weekes et al.\ 2002]{weekes02}
Weekes, T.C., et al. 2002, Astropart.Phys., 17, 221 





\end{thebibliography}
\end{document}